# RF Propagation Analysis of MICAz Mote's Antenna with Ground Effect

Syed Hassan Ahmed[1a], Safdar H. Bouk[1b], Nadeem Javaid[1c], and Iwao Sasase[2]

[1] Department of Electrical Engineering, COMSATS Institute of Information Technology, Islamabad, Pakistan.
[2] Department of Information and Computer Science, Keio University, Japan.
[a] sani9585@gmail.com, {[b] bouk, [c] nadeemjavaid}@comsats.edu.pk, [2] sasase@ics.keio.ac.jp

*Abstract*—**In this paper we analyzed the Radio Frequency (RF) propagation characteristics of monopole antenna in MICAz mote. During the experimental analysis, two scenarios are considered. In Scenario-I, a pair of MICAz nodes (one transmitting and one receiving node) are placed on the ground and the RSSI is measured in presence of the ground effect. In Scenario-II, only the transmitting node is placed above the ground; however, the receiving node is placed on the ground. The RSSI is measured by changing the antenna orientation at different angles and distances between them. The results show that the ground effect, antenna orientation and distance between the sensor nodes drastically affect the RSSI.**

## I. INTRODUCTION

Wireless Sensor Networks (WSNs) [1] have been intensively explored by the research community around the world due to their wide range of applications, cost effectiveness, emergence with other technologies and low maintenance cost. The WSNs are being used in multi-disciplinary areas [2]: healthcare, transportation, agriculture, military etc. Specific sensing modules (with various sensing capabilities i.e. temperature, light, humidity, soil moisture, etc. or combination of these) are used in WSN to sense the specific characteristics.

Recently, research in the field of WSNs has been centered on the deployment issues [3], where the main focus is on the impact of antenna on the WSNs throughput [4]. Several deployment scenarios have been considered in the WSN where different antennas have been attached with the sensor nodes and their effect on the signal strength has been monitored [5]-[8].

The antenna impact in WSNs has been observed by placing pair of nodes, one as a transmitter and the other as a receiver, in varying environments e.g. laboratory, plain open field, field with grass etc. The characteristics of the transmitter sensor node's antenna (here we call it Antenna Under Test or AUT) is observed by placing at different heights, distances to the receiving node, polarizations (either vertical or horizontal) and rotated at 360° in azimuth direction. The antenna of the receiving node (called Measuring Antenna or MA) records the signal strength (specifically the Received Signal Strength Indicator – RSSI) [9] from the packets sent by the transmitting sensor node. The height, location and the angle of the MA remain static during the course of the experiment.

The MICAz node (MPR2400CA) has been widely used in the WSNs research and deployment experiments. In this work we used the MICAz OEM Edition (MPR2600J [10] Japan), MEMSIC Inc [11], previously Crossbow. The MICAz OEM Edition sensor node is equipped with Atmega128L processor, CC2420 [12] RF Chip and a half-wave external monopole antenna. CC2420 is designed for the low-power and low-rate systems that operate at ISM Frequency Band from 2400MHz to 2483.5MHz.

The LQI and RSSI are two important RF link parameters that are measured and provided by CC2420. RSSI of the received signal is computed from the 8 symbol periods (128μs) and stored as a singed 2's complement value in the RSSI_VAL register. RF signal power (P) is computed as:

$$P = RSSI\_VAL + RSSI\_OFFSET \text{ (dBm)} \qquad (1)$$

where RSSI_OFFSET is about -45. The RSSI measurements of RF signal power in this paper are computed as P in the above expression.

The data rate provided by the CC2420 is 250kbps and the typical RF receiving sensitivity level is -94dBm. It indicates that received signal power should be more than the minimum receiving sensitivity level for proper communication. It is also observed in [13] that the radiation pattern of the MICAz Mote's monopole antenna is far more different than its theoretical circular radiation pattern, even in propagation environment similar to the free-space propagation.

In this paper, the RF propagation of MICA's mote's antenna has been analyzed in presence of the ground effect, where AUT and MA are placed *0m* above the ground. Different polarization angles of AUT and distances between AUT and MA are also considered during the analysis. Rest of the paper is organized as follows: Previous work is discussed in Section II. Simulation setup and results are presented in Section III. Finally, Section IV summarizes the paper.

## II. PREVIOUS WORK IN THE AREA OF MICAZ MOTE'S ANTENNA ANALYSIS

MICAz Mote's antenna characteristics have been analyzed several times in the past where different environments with horizontal and vertical orientations have been considered. T.

Camilo et. al. in [5] performed experimental evaluation of radio characteristics of MICAz and ESB motes' antenna at different positions in an anechoic chamber. Experiments were also performed in various indoor and outdoor environments in presence of different factors that influence the radio propagation e.g. WiFi, Bluetooth, Mobile phone, etc. Both sender and receiver nodes were placed 5m apart and 1.5m above the ground.

In [6], authors investigate the antenna radiation pattern of the custom developed prototype of the MICAz mote in the Radio Frequency (RF) Anechoic Chamber. The Anechoic Chamber resembles the free space propagation model because the walls, floor and ceiling of the chamber are mounted with the RF absorber material. The horn antenna is used at the transmitter side and the MICAz prototype at the receiving side. Both are placed 5.8 meters apart and the prototype is raised 1 meter high above the Anechoic Chamber floor. To MICAz prototype is mounted on the mount inside the Anechoic Chamber that rotates horizontally to get the 360° RF pattern. This study does not take into account the ground effect and multipath reflection and fading.

The received signal strength of MICAz and Tmote motes with Dipole, Bi-quad and Yagi antennas are analyzed in [7]. The receiver and transmitter antennas are placed 5.4m apart, 40cm above the ground and in 0° parallel section alignment to measure the propagation characteristics in a laboratory room. The effect of distance on signal strength is measure between two antennas by placing them 80cm above the ground. The signal strength is measured by using the portable spectrum analyzer and through RSSI parameter.

Z. Fang et. al. [8] RSSI of the custom designed sensor node (SKLTT), equipped with CC2420 radio chip with monopole antenna, is analyzed in lab and outdoor environment. The ground effect of radio signal in the experiments is not considered by placing the SKLTT source and destination motes about 36.6cm above the ground.

The previous work in the area of the antenna radio propagation characteristic analysis of wireless sensor nodes is focused on vertical and horizontal polarization. The other polarization angles and ground effect is not considered in the previous work. In this paper, we analyze the RF propagation characteristic of MICAz Mote's antenna with different elevation angles and distances with ground level effect.

### III. EXPERIMENTAL SETUP AND RESULTS

#### A. Experimental Setup

To analyze the ground effect on RF propagation of MICAz mote's antenna radiation pattern, we use a pair of MICAz OEM Edition (MPR2600J) motes in the experiment. The AUT MICAz mote continuously transmits packets and the other mote with MA is connected to the computer and it receives, computes and stores the RSSI based on the received packets. Two experimental scenarios are considered in this experiment. In Scenario-I, two motes are placed on the ground (0cm above the ground) at *3m*, *5m* and *7m* apart. On the other hand, in Scenario-II, MICAz mote with MA is placed on the ground and the MICAz mote with AUT is placed about *0.65m* above the ground on a wooden stool. The MICAz mote with MA and the stool are also placed *3m*, *5m* and *7m* apart in this scenario. In both the scenarios at each distance, the RSSI of the AUT mote is measured by rotating it in anticlockwise azimuth direction by setting AUT's Z-axis at $\theta=0°$, $45°$ and $90°$, as shown in Fig.1. The experimental scenario-I and II are shown in Fig. 2(a) and Fig. 2(b), respectively.

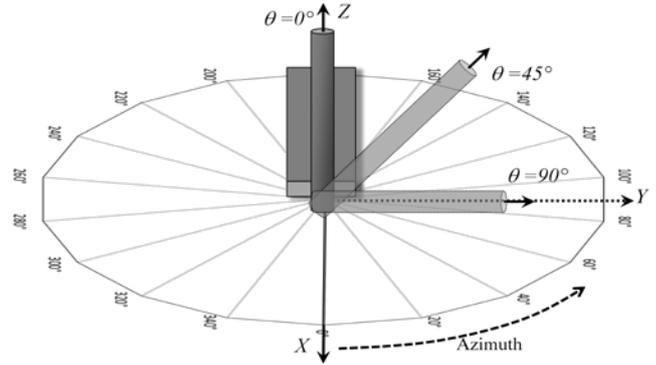

Figure 1. AUT's Z-axis during experimental analysis.

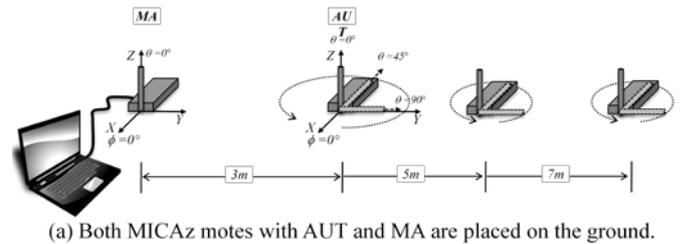

(a) Both MICAz motes with AUT and MA are placed on the ground.

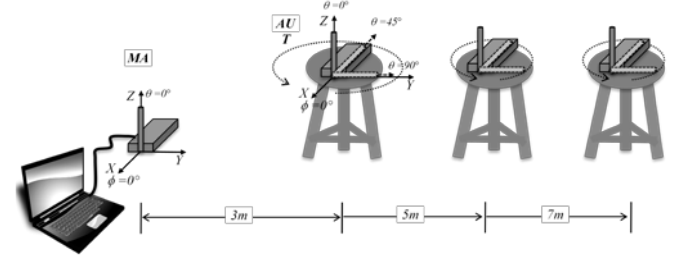

(b) MICAz mote with MA is placed on the ground and AUT MICAz mote is placed on the stool.

Figure 2. Experimental Scenarions (a) Scenario-I (b) Scenario-II

#### B. Experimental Results.

The experiment to analyze the effect of the MPR2600J antenna orientation in presence of the ground effect on the RF propagation pattern, are performed in a Laboratory environment. The RSSI polar graphs for varying distances and orientations in the aforesaid scenarios are discussed below.

Figure 3 to Figure 5 show the RSSI value of MICAz with AUT placed at 3m, 5m and 7m apart, respectively, as in Scenario-I.

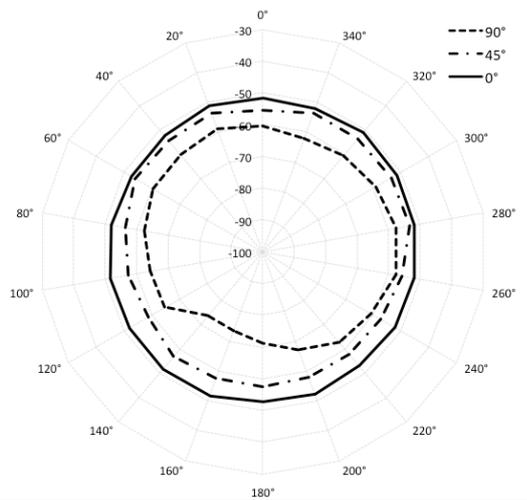

Figure 3.  RF radiation pattern of MICAz with AUT in Scenario-I at 3m apart

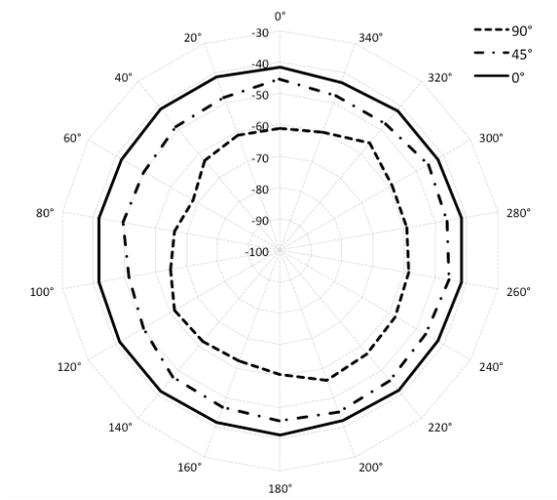

Figure 6.  RF radiation pattern of MICAz with AUT in Scenario-II at *3m* apart

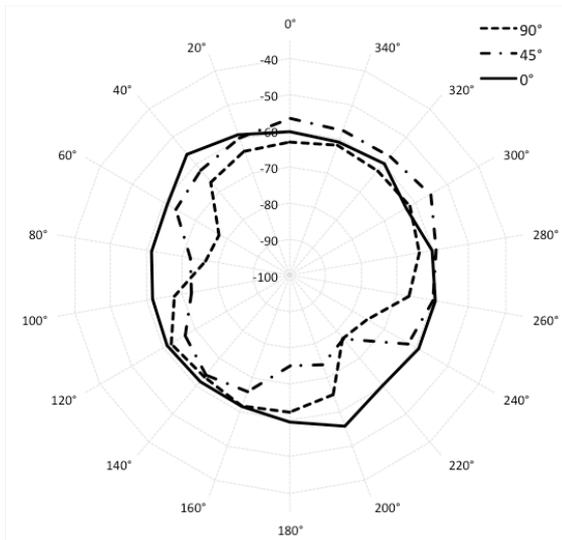

Figure 4.  RF radiation pattern of MICAz with AUT in Scenario-I at 5m apart

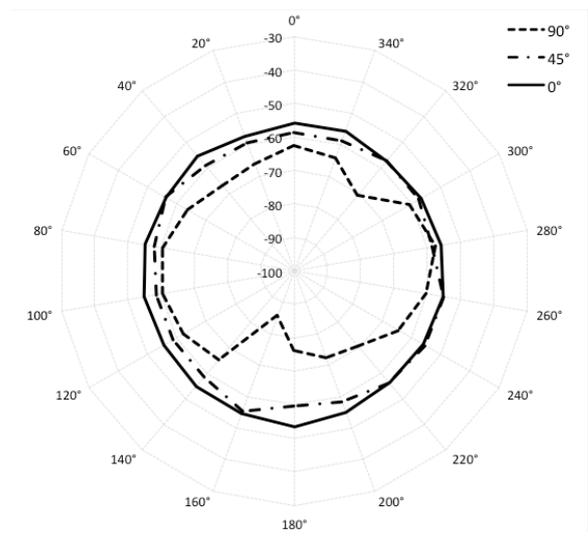

Figure 7.  RF radiation pattern of MICAz with AUT in Scenario-II at *5m* apart

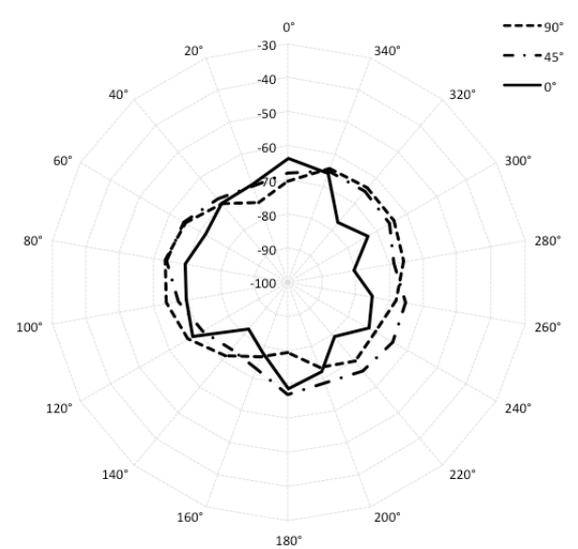

Figure 5.  RF radiation pattern of MICAz with AUT in Scenario-I at 5m apart

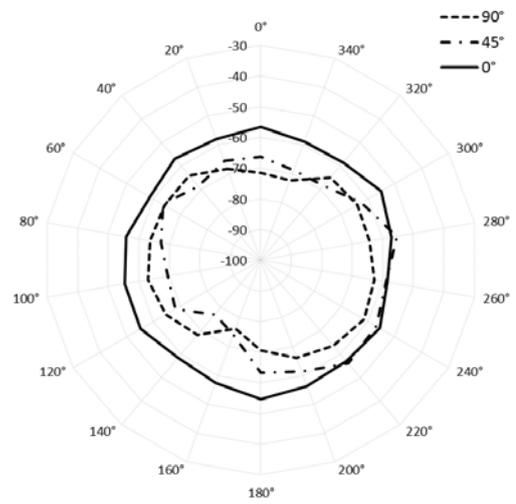

Figure 8.  RF radiation pattern of MICAz with AUT in Scenario-II at 7m apart

It is evident from the results that for $\theta=0°$ has strong RSSI compared to the $\theta=45°$ and $\theta=90°$. As distance increases, the signal strength decreases and even for the $\theta=0°$, RSSI fluctuations increase drastically. RSSI value at main null for different distances and $\theta$, are summarized in the following table.

TABLE I.  MAIN NULL RSSI VALUES OF SCENARIO-I

| Distance Between AUT & MA | Z-axis of AUT | | |
|---|---|---|---|
| | θ=90° | θ=45° | θ=0° |
| 3m | -73 dBm | -58 dBm | -53 dBm |
| 5m | -77 dBm | -77 dBm | -63 dBm |
| 7m | -78 dBm | -74 dBm | -81 dBm |

TABLE II.  RSSI IMPROVEMENT IN SCENARIO-II COMPARED TO SCENARIO-I

| Distance Between AUT & MA | Z-axis of AUT | | |
|---|---|---|---|
| | θ=90° | θ=45° | θ=0° |
| 3m | 2.48 dBm | 8.39 dBm | 10.21 dBm |
| 5m | 0.55 dBm | 6.93 dBm | 4.53 dBm |
| 7m | 2.61 dBm | 2.55 dBm | 15.68 dBm |

The Scenario-II experimental results for varying distances and *Z-axis* are shown in Fig. 6 to Fig. 8. Results show that the RSSI measurements for $\theta=0°$ at varying distances are almost circular and are not affected by ground effect, because AUT mote is placed about 0.65m above the ground. By comparing the results of Scenario-I and Scenario-II we observed that RSSI signal strength has been considerably improved. This improvement is summarized in Table-II.

The standard deviation in RSSI versus varying distances between AUT and MA for different $\theta$ angles in Scenario-I and Scenario-II, are shown in Fig. 9 and Fig. 10, respectively. It is evident from the experimental results that $\theta=0°$ has less deviation compared to the other angles at smaller distance in presence of ground effect. On the other hand, if the ground effect is minimized, as in Scenario-II, $\theta=0°$ has minimum standard deviation compared to 45 and 90 from small to the long distances between AUT and MA.

IV. CONCLUSION

In this paper we performed the experimental analysis of RF propagation characteristics of the MICAz Mote's antenna. It has been observed during the experimental analysis that along with the distance, ground effect drastically decreases the RSSI of an antenna. Experimental results show that the AUT at $\theta=0°$ and placement above the ground level has more stable RSSI compared to the $\theta=45°$ and $\theta=90°$ and its placement on the ground. Therefore, we conclude that the RF antenna propagation of MICAz mote is an important factor in planning the sensor network deployment and optimized WSNs area coverage solutions.

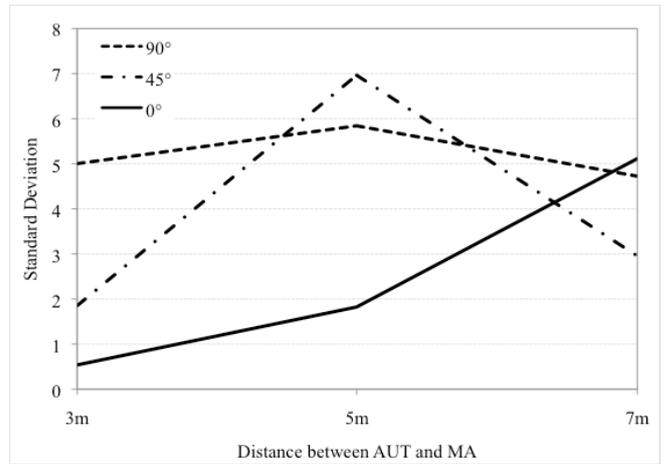

Figure 9. Standard Deviation versus distance between AUT and MA for varying θ angles in Scenario-I.

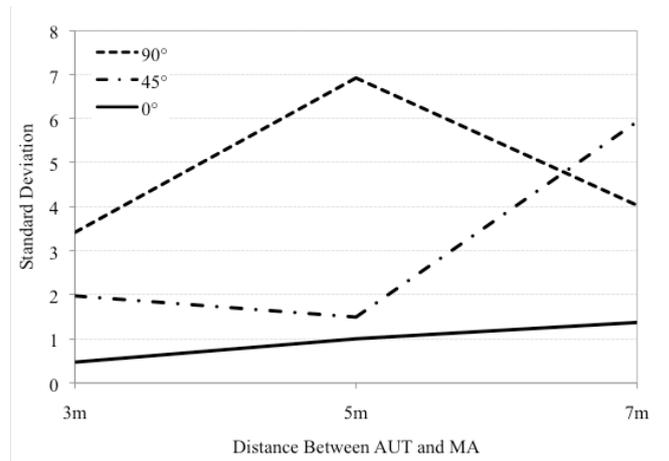

Figure 10. Standard Deviation versus distance between AUT and MA for varying $\theta$ angles in Scenario-II.

ACKNOWLEDGMENT

This work was partly supported by Keio University 21st Century Center of Excellence Program on "Optical and Electronic Device Technology for Access Network" and Fujitsu Laboratories.

REFERENCES

[1] M.A.M. Vieira, C.N. Jr. Coelho, D.C. Jr. da Silva, and J.M. da Mata, "Survey on wireless sensor network devices" In Proceedings of The IEEE Conference on Emerging Technologies and Factory Automation, 2003, (ETFA '03), Lisbon, Portugal, 16-19 Sept. 2003, Vol. 1, pp. 537-544.

[2] A. Alemdar, and M. Ibnkahla, "Wireless sensor networks: Applications and challenges", In Proceedings of the 9th International Symposium on Signal Processing and Its Applications, 2007. ISSPA 2007, Sharjah, United Arab Emirate, 12-15 Feb. 2007, pp. 1 – 6

[3] N. Akshay, M.P. Kumar, B. Harish, S. Dhanorkar, "An efficient approach for sensor deployments in wireless sensor network," In Proceedings of the *International Conference on Emerging Trends in Robotics and Communication Technologies (INTERACT),* 2010, Chennai, India, 3-5 December 2010, pp.350-355.

[4] M. Wadhwa, Song Min, V. Rali, and S. Shetty, "The impact of antenna orientation on wireless sensor network performance",Computer Science and Information Technology, 2009. ICCSIT 2009. 2nd IEEE International Conference on, 8-11 Aug. 2009, pp. 143–147

[5] T. Camilo, P. Melo, A. Rodrigues, L. Pedrosa, J. S. Silva, R. Neves, R. Rocha, F. Boavida; "Wireless Sensor Network Deployment: an Experimental Approach", Book Chapter, in "Wireless Mesh Networking", G. Aggelou, McGraw-Hill, International, 2008, pp. 352~362.

[6] W. Su and M. Alzaghal, "Channel Propagation Measurement and Simulation of MICAz mote" WSEAS Transactions on Computers, Issue 4, Volume 7, April 2008, pp. 259~264.

[7] [jar] Joaquim A. R. Azevedo and F. E. Santos, "Signal Propagation Measurements with Wireless Sensor Nodes", Project INTERREG III B (05/MAC/2.3/C16) - FEDER (2005-2008), July 2007.

[8] [fang] Z. Fang; Z. Zhao, D. Geng, Y. Xuan, L. Du, and X. Cui, "RSSI variability characterization and calibration method in wireless sensor network," in Proceedings of the *IEEE International Conference on Information and Automation (ICIA), 2010*, Beijing, China, 20-23 June 2010, pp.1532-1537.

[9] [2011] K. S. Prabh and J. H. Hauer, "Opportunistic Packet Scheduling in Body Area Networks", In Proc. of 8[th] European Conference on Wireless Sensor Networks (EWSN), Bonn, Germany, February 2011.

[10] [Japan] Crossbow Japan, "IRIS and MICAz: Wireless ZigBee Type Sensor Networks" http://www.xbow.jp/zigbee-smartdust.html

[11] [Memsic] Memsic Inc., "Wireless Modules" http://www.memsic.com/products/wireless-sensor-networks/wireless-modules.html

[12] Texas Instruments, "CC2420 - Single-Chip 2.4 GHz IEEE 802.15.4 Compliant and ZigBee Ready RF Transceiver". http://www.ti.com/product/cc2420

[13] [RSSI_2] E.B.S. Tan, J.G. Lim, W.K.G. Seah, and S.V. Rao, "On the Practical Issues in Hop Count Localization of Sensors in a Multihop Network", In Proceedings of the IEEE 63rd Vehicular Technology Conference, 2006 (VTC 2006-Spring), Melbourne, Australia, 7-10 May 2006, pp. 358-362